\documentclass[conference]{IEEEtran}
\IEEEoverridecommandlockouts

\usepackage{cite}
\usepackage{amsmath,amssymb,amsfonts}
\usepackage{algorithmic}
\usepackage{graphicx}
\usepackage{textcomp}
\usepackage{booktabs}
\usepackage[most]{tcolorbox}
\usepackage{mdframed}
\usepackage{todonotes}
\usepackage{amssymb}
\usepackage{wasysym}
\usepackage{pifont}
\usepackage{balance}

\usepackage{xcolor}
\usepackage{float}
\usepackage{hyperref}
\usepackage{listings}
\usepackage{cleveref}
\usepackage{pifont}
\usepackage{xspace}
\usepackage{tabularx}
\usepackage{enumitem}
\usepackage{arydshln}
\usepackage{makecell}
\usepackage{tikz}
\usepackage{subcaption}
\usepackage{soul}

\DeclareRobustCommand{\cb}[6]{%
  \tikz[baseline={(r.base)}]{
    \node[inner sep=0pt, outer sep=0pt] (r) {};%
    \begin{scope}[yshift=-2pt] 
      \draw[
        fill={rgb,1:red,#1; green,#2; blue,#3},
        draw={rgb,1:red,#4; green,#5; blue,#6},
        line width=0.4pt
      ] (0,0) rectangle (8pt,8pt);
    \end{scope}
  }%
}

\mathchardef\UrlBreakPenalty=1000
\mathchardef\UrlBigBreakPenalty=1000

\newcolumntype{L}[1]{>{\raggedright\arraybackslash}p{#1}}
\newcolumntype{C}[1]{>{\centering\arraybackslash}p{#1}}
\newcolumntype{R}[1]{>{\raggedleft\arraybackslash}p{#1}}

\def\BibTeX{{\rm B\kern-.05em{\sc i\kern-.025em b}\kern-.08em
    T\kern-.1667em\lower.7ex\hbox{E}\kern-.125emX}}

\begin{document}

\title{PPTAM$\eta$: Energy Aware CI/CD Pipeline for Container Based Applications}

 \author{\IEEEauthorblockN{Alessandro Aneggi, Xiaozhou Li,Andrea Janes}
 \IEEEauthorblockA{\textit{Free University of Bozen-Bolzano}\\
 Bolzano, Italy \\
 \{aaneggi, xli, ajanes\}@unibz.it}
 }

\maketitle

\begin{abstract}
Modern container-based microservices evolve through rapid deployment cycles, but CI/CD pipelines still rarely measure energy consumption, even though prior work shows that design patterns, code smells and refactorings affect energy efficiency. We present PPTAM$\eta$, an automated pipeline that integrates power and energy measurement into GitLab CI for containerised API systems, coordinating load generation, container monitoring and hardware power probes to collect comparable metrics at each commit. The pipeline makes energy visible to developers, supports version comparison for test engineers and enables trend analysis for researchers. We evaluate PPTAM$\eta$ on a JWT-authenticated API across four commits, collecting performance and energy metrics and summarising the architecture, measurement methodology and validation.
\end{abstract}

\section{Introduction}

Energy efficiency is critical for sustainable software development and operational costs~\cite{Venters2023}. CI/CD pipelines are widely adopted, but they rarely provide consistent mechanisms to assess energy use. Existing solutions for load generation, system observation and power tracking operate separately, so development teams cannot reliably determine how code changes affect overall energy consumption.

A systematic mapping study by Poy et al.\cite{Poy2025} showed that design patterns, code smells and refactoring decisions substantially affect energy consumption, yet developers lack tools to collect energy data for everyday modifications. PPTAM$\eta$ extends PPTAM\cite{Avritzer2019}, an orchestrator for automated performance tests, with systematic CPU and DRAM power measurements after each commit. A second mapping study~\cite{DOSSANTOS2025107641} identified energy regression testing as an overlooked dimension, which PPTAM$\eta$ addresses. For example, switching a JSON Web Token (JWT) signing algorithm (e.g., from HS256 to RS256) can change energy consumption while leaving throughput almost unaffected.

Recent work has integrated performance testing into DevOps; for instance, Camilli et al.\cite{Camilli2022} presented integrated performance and reliability testing for microservices. However, energy measurement is not systematically included. Kruglov et al.\cite{Kruglov2022} examined energy measurement within CI/CD pipelines, but their approach is tightly coupled to object-oriented metrics and a specific workflow. In contrast, a plugin-based framework like PPTAM decouples measurement concerns, allowing different aspects to be captured independently and combined later during analysis.

We therefore integrate continuous energy measurement into the everyday workflow of software engineers, test practitioners and researchers. PPTAM$\eta$ extends PPTAM with energy measurement within GitLab CI\footnote{\url{https://docs.gitlab.com/ci/}} and addresses three aspects: (1) \emph{integrated measurement}, by coordinating monitoring tools and synchronising data collection at each commit; (2) \emph{methodology grounded in real scenarios}, where test engineers define representative usage scenarios and the pipeline returns comparable metrics for each commit; and (3) \emph{longitudinal analysis}, enabling researchers to collect energy data across version sequences to identify trends and regressions.

Our proof-of-concept applies PPTAM$\eta$ to a JWT-authenticated microservice across four commits, collecting performance and energy metrics and demonstrating the feasibility of integrated measurement within CI/CD workflows.

In the following, we introduce PPTAM$\eta$ (Section\ref{Tool}), describe its architecture (Section\ref{Architecture}), validation (Section\ref{Validation}), discuss related work (Section\ref{Related}) and conclusion (Section\ref{Conclusion}).

\section{Tool overview and envisioned usage}\label{Tool}

\subsection{Envisioned Users and Workflows}

\begin{figure}[htbp]
\centering
\includegraphics[width=\columnwidth]{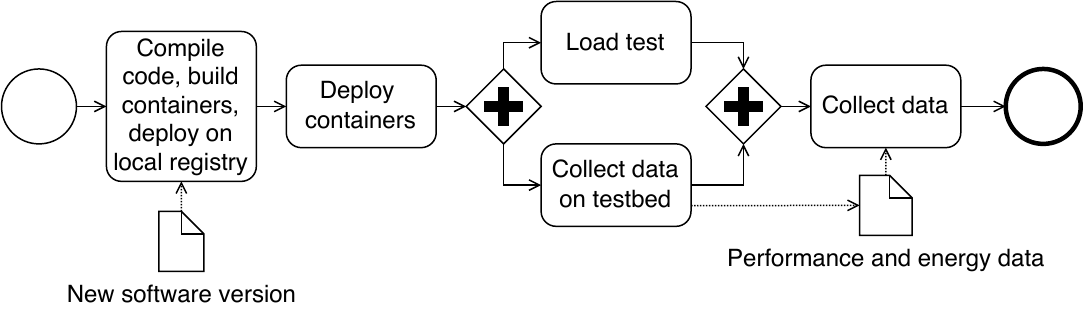}
\caption{Sequence diagram of pipeline process (BPMN notation). }
\label{fig:bpmn}
\end{figure}
PPTAM$\eta$ is intended for three main categories of users: developers, who commit new code and receive feedback from the tool; test engineers, who define the characteristics of the experiments and, in parallel, obtain information on energy consumption; and researchers, who can automatically analyse multiple software versions over time.

The workflow of PPTAM$\eta$, summarized in Figure \ref{fig:bpmn}, begins when a developer commits and pushes changes to the GitLab repository that contains both the system under test and the PPTAM$\eta$ configuration files. The GitLab Runner\footnote{\url{https://gitlab.com/gitlab-org/gitlab-runner/}} then triggers the PPTAM$\eta$ pipeline on a dedicated testbed machine, which builds the updated Docker\footnote{\url{https://www.docker.com}} container images and pushes them to the local container Registry. PPTAM \footnote{\url{https://github.com/pptam/pptam-tool}} 
then orchestrates the deployment of the system under test on the Docker Swarm\footnote{\url{https://docs.docker.com/engine/swarm/}} cluster and launches the Locust\footnote{\url{https://locust.io}} load test for the selected scenario. While the test runs, cAdvisor\footnote{\url{https://github.com/google/cadvisor}}, Perf\footnote{\url{https://web.eece.maine.edu/~vweaver/projects/perf_events/}} and PowerJoular\footnote{\url{https://github.com/joular/powerjoular}} operate under the control of a PPTAM Agent\footnote{\url{https://github.com/pptam/pptam-agent}} to gather container level resource metrics and power measurements. After the experiment completes, PPTAM$\eta$ processes and aggregates the logs, and stores the resulting metrics. Results are then visualized through a dashboard or can be extracted to perform specific studies into, e.g., Jupiter Notebooks\footnote{\url{https://jupyter.org}}.

\subsection{Software engineering challenges and setup}

From a software engineering perspective, PPTAM$\eta$ supports automatic energy-aware regression testing for container-based systems, coordinating heterogeneous tools (load generators, container monitors and RAPL-based power probes) and enabling repeatable, statistically valid experiments within the time and resource constraints of continuous integration. It collects multiple metrics, summarised in Table~\ref{tab:metrics}, produced by the tools listed in Table~\ref{tab:tools}.

Using PPTAM$\eta$ requires the system under test to be containerised: developers provide Dockerfiles and container configurations, as well as a design folder that defines the experiment configuration (for example the Locust load profile and execution type). DevOps engineers or researchers provision the bare-metal testbed and install the required software stack (Python environment and system-level monitoring tools), and configure the \texttt{.gitlab-ci.yml} file so that the PPTAM$\eta$ pipeline is integrated into the continuous integration workflow.

\begin{table}[t]
\setlength{\tabcolsep}{.7pt}
\centering
\caption{Measurements, units, and their classification}
\label{tab:metrics}
\begin{tabular}{@{}l@{\hspace{2pt}}L{3.22cm}L{2cm}ccccc cc@{}}
\toprule
\textbf{Tool} & \textbf{Metric} & \makecell[l]{\textbf{Unit}} & \rotatebox{90}{\textbf{Host\textsuperscript{a}}} & \rotatebox{90}{\textbf{Container\textsuperscript{a}}} & \rotatebox{90}{\textbf{Endpoint\textsuperscript{a}}} & \rotatebox{90}{\textbf{Direct\textsuperscript{b}}} & \rotatebox{90}{\textbf{Derived\textsuperscript{b}}} & \rotatebox{90}{\textbf{Point\textsuperscript{c}}} & \rotatebox{90}{\textbf{Cumulative\textsuperscript{c}}} \\
\midrule
cAdvisor & Memory usage & Byte (b) & & \ding{53} & & \ding{53} & & \ding{53} & \\\hdashline
Powerjoular & CPU power consumption & Watt (W) & & \ding{53} & & \ding{53} & & \ding{53} & \\\hdashline
Powerjoular & CPU utilization & Percent (\%) & \ding{53} & \ding{53} & & \ding{53} & & \ding{53} & \\\hdashline
perf & DRAM power consumption & Microjoule ($\mu$J) & \ding{53} & & & & \ding{53} & & \ding{53} \\\hdashline
locust & Response time & Millisecond (ms) & & & \ding{53} & \ding{53} & & \ding{53} & \\\hdashline
locust & Failures & Amount & & & \ding{53} & \ding{53} & & \ding{53} & \\
\bottomrule
\multicolumn{10}{l}{\textsuperscript{a}~Scope dimension \textsuperscript{b}~Origin dimension \textsuperscript{c}~Temporal dimension}
\end{tabular}
\end{table}

\label{Architecture}

\begin{figure}[htbp]
    \centering
    \includegraphics[width=\columnwidth]{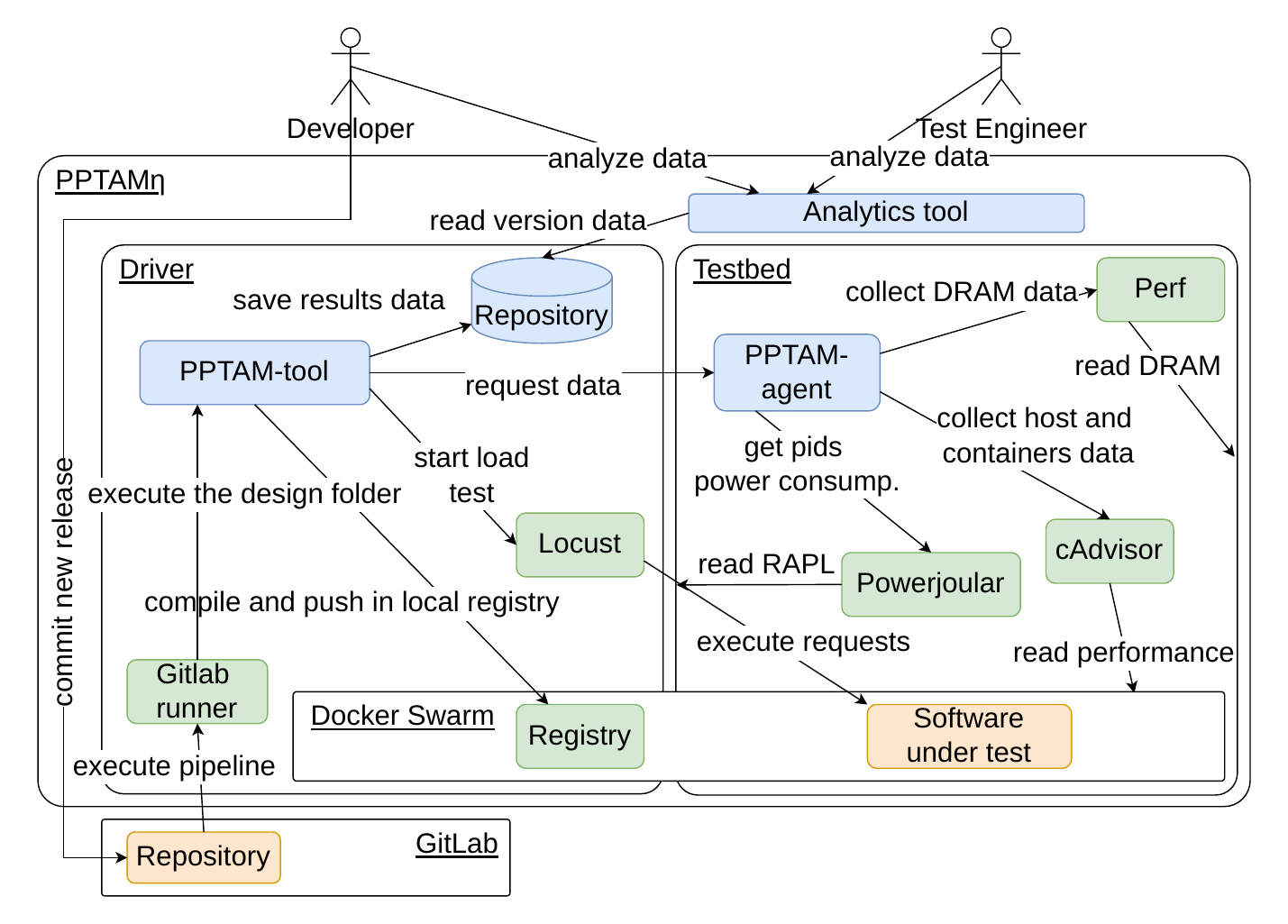}
    \caption{Diagram of PPTAM$\eta$  (C4 notation \cite{brown2026}). \cb{0.863}{0.906}{0.980}{0.451}{0.553}{0.733} blue indicate PPTAM$\eta$ components; \cb{0.847}{0.906}{0.835}{0.549}{0.698}{0.431} green indicate used elements; \cb{0.984}{0.906}{0.812}{0.882}{0.741}{0.404} orange indicate software under test; }
    \label{fig:architecture}
\end{figure}

Figure~\ref{fig:architecture} sketches the main elements of PPTAM$\eta$. The architecture is organized in three logical parts: 1) the PPTAM and GitLab side, where the experiments are defined and orchestrated; 2) the driver on the GitLab runner, which executes the continuous integration pipeline; and 3) the testbed, where the monitoring tools and the system under test run.

PPTAM$\eta$ reuses the plugin-based configuration of PPTAM. New plugins coordinate the activation of monitoring tools and the transfer of data. The system under test is deployed as a set of Docker containers managed by Docker Swarm.

Our implementation requires multiple machines. Specifically, it uses a \textit{driver}, which controls and manages the experiment, including the processing and deployment of the \textit{software under test}, and a \textit{testbed}, which is the host where the SUT is deployed and executed during the load test.

As shown in Figure~\ref{fig:architecture}, the main components deployed on the two machines and their responsibilities are described in the following.

\begin{table}[h]
\centering
\begin{tabular}{@{}L{1.3cm}L{7.1cm}@{}}
\hline
\textbf{Component} & \textbf{Description} \\
\hline

Driver & Hosts the GitLab Runner, PPTAM tool, Locust, the local container registry, and the repository data. \\\hdashline

Testbed & Baremetal machine that stores logs, runs the PPTAM agent, and exposes collected data to the PPTAM tool. \\\hdashline

GitLab Runner & Triggers the PPTAM$\eta$ pipeline, builds and publishes container images, and starts the PPTAM tool with the selected design folder and test plan. \\\hdashline

PPTAM GitLab Repository & Stores the system under test, Dockerfiles, and PPTAM$\eta$ configuration files that define experiment parameters and load profiles. \\\hdashline

PPTAM Tool & Orchestrates experiments based on the test plan, controls plugins and agent, and aggregates structured results. \\\hdashline

Docker Swarm and Registry & Deploys versioned container images from the local registry on the testbed, ensuring a clean and reproducible environment for each run. \\\hdashline

PPTAM Agent & Coordinates monitoring tools on the testbed, manages process to container mappings, and handles start or stop commands for data collection. \\\hdashline

Locust & Generates HTTP workloads and provides performance metrics based on a consistent load profile. \\\hdashline

cAdvisor & Collects container and host resource usage such as CPU, memory and network traffic to support bottleneck analysis. \\\hdashline

Perf & Reads hardware performance counters and DRAM related events, providing low level data relevant to power and memory behaviour. \\\hdashline

PowerJoular & Reads RAPL power measurements and associates them to containers or processes through the PPTAM agent. \\\hdashline

Analytics Tool & Visualises metrics, computes indicators like energy per request, and supports comparison between system versions. \\
\hline
\end{tabular}
\caption{Overview of the PPTAM$\eta$ components}
\label{tab:tools}
\end{table}

\subsection{Methodology implied for users}

The architecture of PPTAM$\eta$ implies a specific methodology for its users: the \textit{test engineer} defines test plans in PPTAM by specifying load levels, durations, number of users, think-time delays, and number of repetitions. The same test plan is applied across different versions of the system under test to preserve comparability. The test engineer also configures which monitoring components must be active and at which sampling interval. The \textit{developer} works mainly at the level of the application and its configuration. When a new version is ready, it is pushed to the repository, the PPTAM$\eta$ pipeline runs automatically, and a report is produced. If a regression is detected, the developer can inspect detailed metric views and refine the code or configuration accordingly.

For each run, the pipeline stores a coherent set of data (see Tab. \ref{tab:metrics}): 1) performance metrics per endpoint and per phase, 2) resource metrics per container, and 3) power and energy metrics per host and per logical group of containers. These metrics allow users to compare different commits and to analyze performance and energy behavior jointly across versions.

\section{Proof-of-concept}\label{Validation}
PPTAM$\eta$ is implemented and integrated with GitLab on a dedicated testbed. The current prototype runs a predefined test plan on a container-based API benchmark and collects performance, resource and power data as described above.

As an illustrative example, PPTAM$\eta$ is applied to a container based API that we designed. The proof-of-concept implements a JWT based login and is evaluated across four successive commits. The evolution starts from an initial version using HS256 encryption with inefficient code, followed by a version that modifies the payload and significantly increases its size, then a version that replaces HS256 with RS256 encryption, and finally a version that optimizes the previously inefficient algorithm. The same load testing workload is used for all tests and consists of multiple login requests with different credentials.

The experiment was conducted using a virtual \textit{driver} and a bare-metal \textit{testbed}, both configured with the required tools (Tab.~\ref{tab:tools}) and connected to a university-hosted GitLab repository. The workload involved ten concurrent users executing three login operations for 100 seconds, with power sampled every second. PPTAM deployed the containers and, after a ten-second warm-up, executed the experiment as specified in \textit{configuration.ini}. The testbed was a Lenovo X1 Yoga 3rd Gen with an Intel Core i5-8250U CPU, which was never fully saturated.

\subsection{Results}
The complete replication package including PPTAM$\eta$, the used proof-of-concept software, and results are available at \cite{tool_package}.
\begin{figure}
    \centering
    \includegraphics[width=\columnwidth]{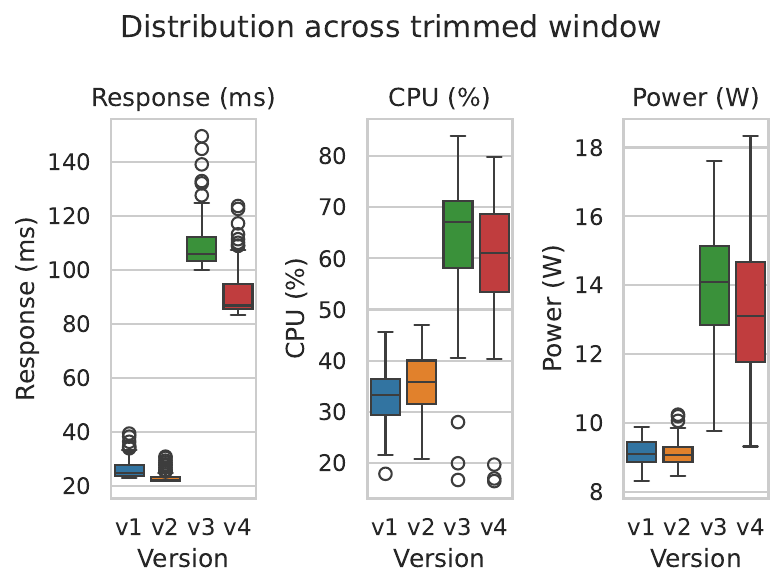}
    \caption{Response time, cpu utilizzation and power consumption of each version tested}
    \label{fig:versions}
\end{figure}
After commits of the four code versions introduced before, PPTAM$\eta$ post-processes the collected metrics and produces a set of graphs and tables in a Jupyter Notebook, which includes, among others: (1) time series of response time and throughput, (2) CPU and memory utilisation per service, (3) CPU package and DRAM power at host level, and derived energy per request.
Figure~\ref{fig:versions} reports boxplots that show how response time, CPU utilisation and power consumption vary across the four committed versions. Figure \ref{fig:versions} shows that versions using HS256 (v1 and v2) and those using RS256 (v3 and v4). The use of RS256 is computationally heavier and leads to higher power usage and CPU utilisation. Between versions v3 and v4 we also observe a decrease in energy consumption due to the introduction of a more efficient search algorithm. In parallel, CPU usage and response time change consistently with the reduction in power consumption.

Table~\ref{tab:final_results} shows that throughput changes only marginally compared to the other metrics, which are more strongly affected by the code changes. An interesting aspect is the comparison between v1 and v2. Although we might expect an increase in power consumption and response time due to the larger payload, the data show a decrease in response time for v2 together with an increase in CPU utilisation and only limited differences in energy and power between these two versions.

Taken together, these visualisations and tables provide a multi-dimensional view of the system that covers several quality aspects, including energy behaviour. They allow users to become more aware of the energy and power consumption of their software, which is the main goal of PPTAM$\eta$.

\begin{table}[htbp]
\centering
\caption{Energy, Latency, and Throughput Metrics per Version}
\label{tab:final_results}
\scriptsize
\setlength{\tabcolsep}{2.6pt} 
\begin{tabular}{lrrccc}
\hline
\textbf{Run} & \textbf{CPU (J)} & \textbf{DRAM (J)} & \textbf{Resp. M(Max) ms} & \textbf{Thr. M(Max)} & \textbf{CPU \% M(Max)} \\ \hline
v1 & 886.75 & 29.71 & 26.5 (39.4) & 12.6 (13.8) & 33.3 (45.6) \\
v2 & 882.70 & 30.01 & 23.1 (30.9) & 12.9 (14.1) & 35.9 (46.9) \\
v3 & 1,381.91 & 29.66 & 109.7 (149.5) & 11.1 (12.1) & 63.1 (83.9) \\
v4 & 1,290.71 & 29.30 & 92.0 (123.7) & 11.3 (12.3) & 58.6 (79.9) \\ \hline
\end{tabular}
\vspace{-1cm}
\end{table}

\subsection{Future work}

The current prototype has been validated on a single benchmark API and requires a careful setup, including a bare-metal testbed and manual configuration of containers, test plans and GitLab CI pipelines. A crucial follow-up question concerns the load profile: the workload should realistically represent system usage to obtain meaningful energy evaluations, yet when new features are added it is unclear whether to keep the original workload for comparability or extend it to exercise new functionality, at the cost of changing the overall load and reducing comparability with earlier versions. We see three main areas to expand this research:
\textit{(1)Extended validation.} Conduct validation studies on representative benchmarks with diverse architectural patterns and deployment scenarios, in order to assess the generalizability of the approach.
\textit{(2)Advanced analytics.} Extend the analytics dashboard with energy-oriented indicators and statistical comparison methods that help engineers distinguish noise from significant energy regressions across versions and new features, and investigate more technology-agnostic ways to express and compare energy efficiency.
\textit{(3)Scalability and tooling.} Improve automation of setup and configuration to reduce manual effort, and explore extensions to other continuous integration platforms beyond GitLab, so that the approach can be adopted more easily in different DevOps environments.

\section{Related work}\label{Related}

Software energy efficiency is increasingly important for reducing environmental impact and operational costs\cite{Poy2025}. PPTAM\cite{Avritzer2019} provides automated infrastructure for production performance testing with dashboard visualisation, while\cite{Camilli2022} integrates testing into DevOps; PPTAM$\eta$ builds on these ideas by adding systematic energy measurement. Load testing frameworks such as Locust and JMeter support scalable experiments, but typical CI/CD pipelines still focus on classical metrics (response time, throughput) and rarely integrate power and energy measurements.

Intel RAPL offers hardware-based power measurement at CPU package and DRAM levels. Hähnel et al.\cite{Hahnel2012} showed that RAPL enables fine-grained energy measurement with low error, and recent validation on modern architectures\cite{Alt2024} confirms its reliability under load. Tools such as Scaphandre\footnote{https://github.com/hubblo-org/scaphandre} and PowerJoular use RAPL to estimate power at process and container level, while PowerAPI\cite{guaman2025containerizing} supports reproducible energy measurement across heterogeneous applications and environments.

Poy et al.~\cite{Poy2025} presented a systematic mapping study on how design patterns, code smells and refactorings affect energy consumption, showing that most code smell removal improves energy efficiency, that pattern effects depend on context and language, and that hardware-based instruments are generally more accurate than software-based tools. Green Metrics Tool\footnote{https://www.green-coding.io/} and similar solutions collect multiple metrics at different granularities, and individual components exist for load generation (Locust, JMeter), energy measurement (Scaphandre, PowerJoular) and container monitoring (cAdvisor). However, an integrated, CI/CD-oriented solution that systematically aligns these components for energy-aware regression testing of container-based systems is still missing.

\section{Conclusion}\label{Conclusion}

This paper presented PPTAM$\eta$, an automated, energy-aware assessment pipeline that integrates power and energy measurement into continuous integration workflows for container-based API systems. By orchestrating load generation, container resource monitoring and hardware-level power measurement in a unified GitLab-based environment, PPTAM$\eta$ enables DevOps and performance engineers to obtain integrated feedback on performance, service quality and energy consumption as their systems evolve.

In summary, PPTAM$\eta$ integrates GitLab CI, Docker Swarm and monitoring tools in a single pipeline, provides a simple methodology to collect comparable performance and energy metrics for each commit, and visualises energy regressions in a JWT-based case study. By making energy consumption visible and measurable at each commit, it supports early detection of energy regressions and helps practitioners reason about the impact of design and implementation choices on the energy footprint of container-based API systems.

\bibliographystyle{IEEEtran}
\bibliography{bib}

\end{document}